\documentstyle[11pt]{article}
\newcommand{\pul}{\textstyle{\frac{1}{2}}}
\newcommand{\la}{\frac{\Lambda}{6}}
\newcommand{\G}{\Gamma^}
\newcommand{\e}{{\bf e}}
\newcommand{\m}{{\bf m}}
\newcommand{\bl}{{\bf l}}
\newcommand{\k}{{\bf k}}
\newcommand{\x}{{\bf x}}
\newcommand{\uu}{{\bf u}}

\newcommand{\Z}{{\bf Z}}

\newcommand{\Aa}{{\rm A}}
\newcommand{\BB}{{\rm B}}
\newcommand{\R}{{\cal R}}
\newcommand{\I}{{\cal I}}
\newcommand{\GG}{{\cal G}}
\newcommand{\A}{{\cal A}}
\newcommand{\B}{{\cal B}}
\newcommand{\C}{{\cal C}}
\newcommand{\LL}{{\cal L}}
\newcommand{\M}{{\cal M}}
\newcommand{\bomega}{\omega\!\!\!\!\hbox{\hskip 0.2 mm}\omega\!\!\!\!\hbox{\hskip 0.2 mm}\omega}

\begin{document}
\large

 {\LARGE{\bf
 \centerline {Gravitational waves in vacuum spacetimes}
 \centerline {with cosmological constant.}
 \centerline {II. Deviation of geodesics and interpretation}
 \centerline {of non-twisting type {\it N} solutions}}}
 \vspace{10mm}
 Ji\v r\'\i\  Bi\v c\' ak  and  Ji\v r\' \i\  Podolsk\' y
 \vspace{5mm}

 {\it
 Department of Theoretical Physics, Faculty of Mathematics and Physics,

 Charles University, V Hole\v sovi\v ck\'ach 2, 180 00 Prague 8, Czech Republic}
 \vspace{5mm}

 {\footnotesize
 Electronic addresses:
        bicak@mbox.troja.mff.cuni.cz,
        podolsky@mbox.troja.mff.cuni.cz}
 \vspace{10mm}

\begin{abstract}
In a suitably chosen essentially unique frame tied to a given observer
in a general spacetime, the equation of geodesic deviation can be
decomposed into a sum of terms describing specific effects:
isotropic (background) motions associated with the cosmological constant,
transverse motions corresponding to the effects of gravitational waves,
longitudinal motions, and Coulomb-type effects.
Conditions under which the frame is parallelly transported along a geodesic
are discussed. Suitable coordinates are introduced and an explicit
coordinate form of the frame is determined for spacetimes admitting
a non-twisting null congruence.
Specific properties of all non-twisting type~{\it N} vacuum solutions
with cosmological constant $\Lambda$ (non-expanding Kundt class
and  expanding Robinson-Trautman class) are then analyzed.
It is demonstrated that these spacetimes can be understood as exact
transverse gravitational waves of two polarization modes ``+'' and
``$\times$'', shifted by ${\pi\over4}$, which propagate ``on''
Minkowski, de Sitter, or anti-de Sitter backgrounds. It is also shown that
the solutions with $\Lambda>0$ may serve as exact demonstrations
of the cosmic ``no-hair'' conjecture in radiative spacetimes with
no symmetry.

\vspace{3mm}
\noindent
PACS number(s): 04.30.-w, 04.20.Jb, 98.80.Hw
\end{abstract}

\newpage
\noindent
{\bf I. INTRODUCTION AND SUMMARY}

\bigskip
In the preceding paper \cite{BPI} we classified non-twisting
type~{\it N} solutions of the vacuum Einstein's equations with
a non-vanishing cosmological constant $\Lambda$ and analyzed
their geometrical properties. Here we wish to discuss their
physical properties. We shall show that these solutions
can be interpreted as gravitational waves propagating in
spacetimes of constant curvature --- in Minkowski, de~Sitter, or
anti-de~Sitter spaces. Our treatment focuses on the analysis of
the equation of geodesic deviation.

We first discuss the equation of geodesic deviation in general
spacetimes (Section II), briefly reviewing and extending
\cite{Pir1}-\cite{Sze} by
using both a Newman-Penrose null tetrad and a physical frame of four
independent vectors $\{ \e_{(a)}\}$ tied to the geodesic with respect
to which the relative motion is studied. In type~{\it N}
solutions only the Newman-Penrose scalar $\Psi_4$ is
non-vanishing.

Starting from Section III we study non-twisting type~{\it N}
solutions with $\Lambda$. As shown in \cite{BPI},
 they comprise the non-expanding Kundt class
$KN(\Lambda)$ and the expanding Robinson-Trautman class
$RTN(\Lambda, \epsilon)$. By analyzing the geodesic
deviation in these spacetimes we demonstrate that they can be
interpreted as exact transverse gravitational waves with two
polarization modes (shifted by $\frac{\pi}{4}$) propagating
``on'' Minkowski, de~Sitter, or anti-de~Sitter space
(depending on values of $\Lambda$).
In the Appendix we calculate exact forms of wave amplitudes.

At the end of Section IV we discuss, for $\Lambda>0$, special
timelike geodesics explicitly.
We demonstrate that observers moving along these geodesics see
waves decaying exponentially fast and the spacetimes to
approach locally the de~Sitter space --- in agreement with
the cosmic no-hair conjecture (see e.g. \cite{Mae}, and
references therein). As in our previous work \cite{BP1}, \cite{BP2},
this is an explicit demonstration of the conjecture under
the presence of waves within exact theory.


\bigskip
\bigskip
\noindent
{\bf II. THE RELATIVE MOTION OF FREE PARTICLES IN A
GENERAL SPACETIME}

\bigskip
It is natural to base the local characterization of radiative
spacetimes on the equation of geodesic deviation
\cite{Pir1}-\cite{Sze}
\begin{equation}
\frac{D^2Z^\mu}{d\tau^2}=
-R^\mu_{\ \alpha\beta\gamma}u^\alpha Z^\beta u^\gamma\ , \label{E2.3}
\end{equation}
where $\uu={{d\x}/{d \tau }}$, $\uu\cdot\uu=-1$,
is the four-velocity of a free test particle (observer),
$\tau$ is the proper time, and $\Z (\tau)$ is the displacement
vector. In order to obtain  invariant results one
sets up a frame $\{ \e_{(a)} \}$ along the geodesic. The frame components
$Z^{(a)}(\tau)$, $\Z=Z^{(a)} \e_{(a)}$, are  invariant quantities.
Choosing $\e_{(0)}=\uu$ and perpendicular space-like unit
vectors $\{\e_{(1)},\e_{(2)},\e_{(3)}\}$ in the local hypersurface
orthogonal to $\uu$, we have
${\bf e}_{(a)} \cdot {\bf e}_{(b)} \equiv g_{\alpha\beta}
e_{(a)}^\alpha e_{(b)}^\beta
=\eta_{(a)(b)}=\hbox{diag}(-1,1,1,1)$. The dual basis is
$\e^{(0)}=-\uu$ and  $\e^{(i)}=\e_{(i)}\ ,\ i=1,2,3$.
By projecting (\ref{E2.3}) onto the frame we get
\begin{equation}
\ddot Z^{(i)}=-R^{(i)}_{\ (0)(j)(0)}Z^{(j)}\ , \label{E2.5a}
\end{equation}
where $Z^{(j)}=\e^{(j)}\cdot\Z=e^{(j)}_\mu Z^\mu$ determine directly the
distance between close test particles,
\begin{equation}
\ddot Z^{(i)}\equiv\e^{(i)}\cdot{{D^2\Z}\over{d\tau^2}}
  =e^{(i)}_\mu{{D^2Z^\mu}\over{d\tau^2}} \label{E2.5b}
\end{equation}
are physical relative accelerations, and
$R_{(i)(0)(j)(0)}=e^\alpha_{(i)}u^\beta e^\gamma_{(j)}
 u^\delta R_{\alpha\beta\gamma\delta}$.
Eq. (\ref{E2.3}) also implies
${d^2Z^{(0)}}/{d\tau^2} = -u_\mu{D^2Z^\mu}/{d\tau^2} =
 R_{\mu\alpha\beta\gamma} u^\mu u^\alpha Z^\beta u^\gamma =0$
so that $Z^{(0)}=a_0\tau+b_0$, $a_0, b_0$ are constants.
Setting $Z^{(0)}=0$, all test particles are ``synchronized''
by $\tau$ (they always stay in the same local hypersurface).
From the definition of the Weyl tensor we get
$R_{(i)(0)(j)(0)}=C_{(i)(0)(j)(0)}+{1\over2}(\delta_{ij}R_{(0)(0)}-
R_{(i)(j)})+{1\over6}R\delta_{ij}$.
Using Einstein's equations,
\begin{equation}
R_{(i)(0)(j)(0)}=C_{(i)(0)(j)(0)}-{\Lambda\over3}\delta_{ij}-
      {\kappa\over2}\left[T_{(i)(j)}-\delta_{ij}(T_{(0)(0)}+
      {2\over3}T)\right]
\ , \label{E2.8}
\end{equation}
$T=T^{(a)}_{(a)}$. Following \cite{KSMH} we introduce null complex tetrad
$\{ \e_{\hat a} \}=\{\e_{\hat 1},\e_{\hat 2},\e_{\hat 3},\e_{\hat 4}\}=
      \{\m,\bar{\m},\bl,\k\}$,
\begin{eqnarray}
&&\m    ={\textstyle{1\over\sqrt 2}}\left(\e_{(1)}+i \e_{(2)}\right)
                        ,\
\bar{\m}={\textstyle{1\over\sqrt 2}}\left(\e_{(1)}-i \e_{(2)}\right)
                        \ ,\label{E2.9}\\
&&\bl ={\textstyle{1\over\sqrt 2}}\left(\uu-\e_{(3)}\right)
                        ,\
\k  ={\textstyle{1\over\sqrt 2}}\left(\uu+\e_{(3)}\right)\ .\nonumber
\end{eqnarray}
Null tetrad components of the Weyl tensor are
(see e.g. \cite{KSMH}, \cite{NP})
\begin{eqnarray}
&&\Psi_0=C_{\alpha\beta\gamma\delta}k^\alpha m ^\beta k^\gamma
        m^\delta  \ ,\qquad
  \Psi_1=C_{\alpha\beta\gamma\delta}k^\alpha l ^\beta k^\gamma
        m^\delta  \ ,\nonumber\\
&&\Psi_2={\pul}C_{\alpha\beta\gamma\delta}k^\alpha l ^\beta
  (k^\gamma l^\delta - m^\gamma \bar{m}^\delta)
                                   \ ,\label{E2.10}\\
&&\Psi_3=C_{\alpha\beta\gamma\delta}l^\alpha k ^\beta l^\gamma
        \bar{m}^\delta  \ ,\qquad
  \Psi_4=C_{\alpha\beta\gamma\delta}l^\alpha \bar{m}^\beta
        l^\gamma \bar{m}^\delta\ .\nonumber
\end{eqnarray}
Regarding expressions (\ref{E2.10}) and inverting relations
(\ref{E2.9}) we obtain
\begin{eqnarray}
&&C_{(1)(0)(1)(0)}= {\pul}\R e\Psi_0
  +{\pul}\R e\Psi_4 -\R e\Psi_2  \ ,\quad
C_{(2)(0)(2)(0)}=-{\pul}\R e\Psi_0
  -{\pul}\R e\Psi_4 -\R e\Psi_2  \ ,\nonumber\\
&&C_{(1)(0)(2)(0)}={\pul}\I m\Psi_0 -{\pul}\I m\Psi_4
\ ,\quad C_{(3)(0)(3)(0)}=2\,\R e\Psi_2  \ ,\label{E2.11}\\
&&C_{(1)(0)(3)(0)}=-\R e\Psi_1\> +\R e\Psi_3  \ ,\quad
C_{(2)(0)(3)(0)}=-\I m\Psi_1 -\I m\Psi_3    \ .\nonumber
\end{eqnarray}
Substituting Eqs. (\ref{E2.8}) and (\ref{E2.11}) into
Eq. (\ref{E2.5a}) we arrive at:
\begin{eqnarray}
\ddot Z^{(1)}&=&{\Lambda\over3}Z^{(1)}\ \ -
               \ \ {\kappa\over2}(T_{(0)(0)}+{2\over3}T)Z^{(1)}+
               {\kappa\over2}T_{(1)(j)}Z^{(j)}+\GG_1 \ ,\nonumber\\
\ddot Z^{(2)}&=&{\Lambda\over3}Z^{(2)}\ \ -
               \ \ {\kappa\over2}(T_{(0)(0)}+{2\over3}T)Z^{(2)}+
               {\kappa\over2}T_{(2)(j)}Z^{(j)}+\GG_2
                        \ ,\label{E2.12a}\\
\ddot Z^{(3)}&=&{\Lambda\over3}Z^{(3)}\ \ -
               \ \ {\kappa\over2}(T_{(0)(0)}+{2\over3}T)Z^{(3)}+
               {\kappa\over2}T_{(3)(j)}Z^{(j)}+\GG_3
               \ ,\nonumber
\end{eqnarray}
where
\begin{eqnarray}
&&\GG_1\equiv +\ \C Z^{(1)} -(\LL_1-\M_1)Z^{(3)}\qquad\qquad\qquad\qquad
                -(\A_++\B_+)Z^{(1)}+(\A_\times-\B_\times)Z^{(2)}
 \ ,\nonumber\\
&&\GG_2\equiv +\ \C Z^{(2)}\qquad\qquad\qquad\qquad+(\LL_2+\M_2)Z^{(3)}
                +(\A_++\B_+)Z^{(2)}+(\A_\times-\B_\times)Z^{(1)}
 \ ,\nonumber\\
&&\GG_3\equiv -2 \C Z^{(3)}  -(\LL_1-\M_1)Z^{(1)}+(\LL_2+\M_2)Z^{(2)}
 \ ,\nonumber
\end{eqnarray}
and
\begin{eqnarray}
&&\C=\R e\Psi_2\ ,\quad\LL_1=\R e\Psi_3\ ,\quad \LL_2=\I m\Psi_3\ ,\quad
  \M_1=\R e\Psi_1\ ,\quad \M_2=\I m\Psi_1\ ,\label{E2.12b}\\
&&\A_+={\pul}\R e\Psi_4\ ,\quad \A_\times={\pul}\I m\Psi_4
  \ ,\quad
  \B_+={\pul}\R e\Psi_0\ ,\quad \B_\times\>={\pul}\I m\Psi_0
  \ .\nonumber
\end{eqnarray}
Equations (\ref{E2.12a}) are well suited for physical
interpretation. The relative motions depend on:
1. the cosmological constant $\Lambda$ responsible for overall background
   isotropic  motions;
2. the energy-momentum tensor $T_{(a)(b)}$ terms describing interaction with
   matter-content;
3. the terms depending on the local free gravitational field,
   and consisting of Coulomb, longitudinal and transverse (outgoing/ingoing)
   components with amplitudes given by $\Psi_A$'s.
In the following we put $T_{(a)(b)}=0$.
Individual terms in Eq. (\ref{E2.12a}) can be interpreted as follows:
\begin{description}
    \item {\bf$\Lambda$-term:}
Assuming $T_{(a)(b)}=0=\Psi_A$,  Eq. (\ref{E2.12a}) reduces to
\begin{equation}
\ddot Z^{(i)}={\Lambda\over3}Z^{(i)}\ .\label{E2.15}
\end{equation}
Considering a sphere of test particles each having a position vector
$\Z$, Eq. (\ref{E2.15}) implies that
the acceleration  of each particle is in the  direction $\Z$ and
has the same magnitude.
Assume the frame $\{\e_{(i)}\}$ to be parallelly transported so that
$\ddot Z^{(i)}= {{D^2(\e^{(i)}\cdot\Z)}/{d\tau^2}}={{d^2Z^{(i)}}/{d\tau^2}}$.
Eqs. (\ref{E2.15}) have solutions
\begin{eqnarray}
Z^{(i)}(\tau)&=&A_i\,\tau+B_i\hskip58mm\hbox{for } \Lambda=0
    \ ,\nonumber\\
Z^{(i)}(\tau)&=&A_i\exp(\sqrt{\Lambda/3}\,\tau)
  +B_i\exp(-\sqrt{\Lambda/3}\,\tau)\hskip10mm\hbox{for } \Lambda>0
    \ ,\label{E2.16c} \\
Z^{(i)}(\tau)&=&A_i\cos(\sqrt{-\Lambda/3}\,\tau)
  +B_i\sin(\sqrt{-\Lambda/3}\,\tau)\hskip9mm\hbox{for } \Lambda<0
    \ ,\nonumber
\end{eqnarray}
where $A_i, B_i$ are constants.
As expected, conformally flat ($\Psi_A=0$) vacuum backgrounds
(Minkowski, de Sitter or anti-de Sitter)
are homogeneous and isotropic, so that the relative motion of
test particles is isotropic.
     \item {\bf$\Psi_4$-term:}
Assuming $\Lambda=0=T_{(a)(b)}$ and $\Psi_0=\Psi_1=\Psi_2=\Psi_3=0$,
Eq. (\ref{E2.12a}) reduces to
\begin{eqnarray}
\ddot Z^{(1)}&=&-\A_+ Z^{(1)}+\A_\times Z^{(2)}   \ ,\nonumber\\
\ddot Z^{(2)}&=&\hskip3.5mm\A_+Z^{(2)}+\A_\times Z^{(1)}\ ,\label{E2.17}\\
\ddot Z^{(3)}&=&0 \ ,\nonumber
\end{eqnarray}
which describe the influence of ``+''  and  ``$\times$''
polarization modes of a transverse gravitational wave with
amplitudes $\A_+$ and $\A_\times$.
If particles, initially at rest, lie in the $(\e_{(1)},\e_{(2)})$ plane,
there is no motion in the longitudinal direction of $\e_{(3)}$.
The ring of particles is deformed into an ellipse, the axes of different
polarizations are shifted one with respect
to the other by ${\pi\over4}$ (such behavior is typical for
linearized gravitational waves --- cf. e.g. \cite{MTW}).
Making a rotation in the transverse plane by an angle $\vartheta$,
\begin{equation}
\e'_{(1)}= \cos \vartheta\, \e_{(1)}+\sin \vartheta\, \e_{(2)}
  \ ,\quad
\e'_{(2)}= -\sin \vartheta\, \e_{(1)}+\cos \vartheta\, \e_{(2)}
 \label{E2.18}
\end{equation}
--- which corresponds to $\m '=\e^{-i\vartheta} \m$ ---
and using Eqs. (\ref{E2.10}) and  (\ref{E2.12b}), we find
\begin{equation}
\A' _+(\tau)= \cos 2\vartheta \A_+ -\sin 2\vartheta \A_\times
  \ ,\quad
\A' _\times(\tau)= \sin 2\vartheta \A_+ +\cos 2\vartheta \A_\times
  \ .\label{E2.19}
\end{equation}
Taking $\vartheta=\vartheta_+(\tau)=-\pul \hbox{Arg}\, \Psi_4$,
then $\A '_+={1\over2}|\Psi_4|$, $\A '_\times=0$ ---
the wave is purely ``+'' polarized for an observer using
$\m_+=e^{-i\vartheta_+}\m$; if
$\vartheta=\vartheta_\times(\tau)=
\vartheta_++{\pi\over4}$, then $\A '_+=0$, $\A '_\times=\pul |\Psi_4|$
--- the wave is purely ``$\times$''  polarized for an observer using
$\m_\times=e^{-i\vartheta_\times}\m$.  The amplitude $\A = \pul|\Psi_4|$
is invariant under the rotation.
A general observer sees a superposition of the two polarization
modes shifted by $\pi\over4$.
\end{description}
One can similarly show \cite{Sze}, \cite{KSMH} that
$\Psi_3$ and $\Psi_2$ terms describe longitudinal modes and
Coulomb-type effects; $\Psi_1$ and $\Psi_0$ terms are equivalent
to $\Psi_3$ and $\Psi_4$ terms (if $\k\leftrightarrow\bl$).

For given principal null vector $\k$ and observer's $\uu$ we have chosen the
frame vector $\e_{(3)}$ according to Eq. (\ref{E2.9}), which
implies  $k_{(1)}=0=k_{(2)}$, $k_{(3)}\not =0$, and  makes
the physical interpretation based on Eq. (\ref{E2.12a}) simpler.
This leads to essentially unique $\k$ and  $\bl$. More precisely,
we easily show the following

\bigskip

{\it Proposition 1}:
Let $\uu$ be the four-velocity $(\uu\cdot\uu=-1)$ and $\k$ be the
null vector. Then there exists a unit space-like vector $\e_{(3)}$ which
is the projection of the null direction given by $\k$ into the
hypersurface orthogonal to $\uu$. Such $\e_{(3)}$ is unique
(up to reflections $\e_{(3)}\rightarrow -\e_{(3)}$) and is given by
$\ \e_{(3)}= -\uu+\sqrt{2}\;\k$, where $\k$ satisfies
$\ \k\cdot\uu=-1/\sqrt{2}$. Another null vector $\bl$ in the plane
($\uu, \e_{(3)}$) such that $\bl\cdot\k=-1$
is then given by $\ \bl\equiv\sqrt{2}\;\uu-\k$. The only remaining
freedom are rotations  in the plane
($\e_{(1)}, \e_{(2)}$) perpendicular to $\e_{(3)}$.

\bigskip
In what follows the orthonormal tetrad and the corresponding
null frame (\ref{E2.9}) determined according to Proposition 1 will
always be assumed.

Notice that Eqs. (\ref{E2.12a}) represent possible motions seen
by an observer with given $\uu$. By making the Lorentz boosts to
other observers with $\uu'$, $\Psi_A$ change (see e.g. \cite{KSMH}).
Thus, the ``strength of gravitational field'' is strongly observer
dependent (cf. \cite{Pir1}, \cite{Mash} and point 5. in the
discussion following Eq. (\ref{E6.12})).

\bigskip
\bigskip
\noindent
{\bf III. THE CHOICE OF COORDINATES AND PARALLELLY
PROPAGATED FRAMES}

\bigskip
We shall now express our frames in coordinates suitable for spacetimes
admitting a {\it non-twisting} null congruence and give the conditions
for the frames to be parallelly transported.
The field $\k$ is orthogonal to null hypersurfaces,
say $u=$const., so that $k^\mu = g^{\mu\nu}u_{,\nu}$.
It is convenient (cf. \cite{NP}, \cite{RT}) to  choose
as coordinates $u=x^3$, parameter $v=x^0$ along the null geodesics
generated by $k^\mu$, and two complex space-like coordinates
$\xi=x^1$ and $\bar\xi=x^2$ that label the geodesics on each surface
$u=$const. The metric then takes the form
\begin{equation}
g_{\mu\nu}=\left(
\begin{array}{cccc}
0&0&0&g_{03} \\
0&g_{11}&g_{12}&g_{13} \\
0&g_{12}&g_{22}&g_{23} \\
g_{03}&g_{13}&g_{23}&g_{33}
\end{array}\right)\ ,\label{E2.40}
\end{equation}
where $g_{22}=\overline{g_{11}}$, $g_{23}=\overline{g_{13}}$ since
$x^2=\overline{x^1}$; all other components are real, and
\begin{equation}
g_{12}>0\ ,\qquad\qquad
D=g^2_{12}-g_{11}g_{22}>0\ , \label{E2.41}
\end{equation}
since the subspace $(\xi,\bar\xi)$ is space-like.
The vector $\k$ is simply
\begin{equation}
k^\mu=(k^0,0,0,0)\ ,\label{E42c}
\end{equation}
and the four-velocity $\uu$ of a particle moving
along a geodesic $x^\mu(\tau)$, is given by
$u^\mu=(\dot v,\dot\xi,\dot{\bar\xi}, \dot u)$, where the dot is
$d/d\tau$ and $\dot u\not=0$ (otherwise the geodesic would not be
time-like).

\bigskip

{\it Proposition 2}:
In coordinates $(v,\xi,\bar\xi,u)$
the interpretation null tetrad introduced in Proposition 1 has the form
\begin{eqnarray}
&&m^\mu=\Bigg( {1\over{g_{03}\dot u}}\left[(g_{12}\dot\xi+g_{22}\dot{\bar\xi}
  +g_{23}\dot u) g_+ -(g_{11}\dot\xi+g_{12}\dot{\bar\xi}+g_{13}\dot u) g_-
   \exp (-i \hbox{\rm Arg}\> g_{11})\right],  \nonumber\\
 &&\hskip50mm
  g_- \exp (-i \hbox{\rm Arg}\> g_{11}) , -g_+, 0\Bigg)\ ,\nonumber\\
&&{\bar m}^\mu=\Bigg({1\over{g_{03}\dot u}}\left[(g_{11}\dot\xi+g_{12}\dot{\bar\xi}
  +g_{13}\dot u) g_+ -(g_{12}\dot\xi+g_{22}\dot{\bar\xi}+g_{23}\dot u) g_-
   \exp (\>i\hbox{\rm Arg}\> g_{11})\right],   \nonumber\\
 &&\hskip50mm
  -g_+, g_- \exp(\>i \hbox{\rm Arg}\> g_{11}), 0\Bigg)  \ ,\label{E2.45}\\
&&l^\mu=\left( \sqrt 2 \dot v+{1\over{\sqrt 2}}{1\over{g_{03}\dot u}},
  \sqrt 2 \dot\xi , \sqrt 2 \dot{\bar\xi} ,  \sqrt 2 \dot u \right)
  \ ,\nonumber\\
&&k^\mu=\left( -{1\over{\sqrt 2}}{1\over{g_{03}\dot u}}, 0, 0, 0 \right)
 \ ,\nonumber
\end{eqnarray}
where $g_\pm=\sqrt{(g_{12}\pm\sqrt D )/(2D)}$.
The tetrad is unique up to trivial reflections and rotations
$m^\mu\rightarrow m^\mu e^{i\vartheta}$.
The corresponding orthonormal frame obtained from Eq.
$(\ref{E2.45})$ using Eq. $(\ref{E2.9})$ is
\begin{eqnarray}
&&e^\mu_{(0)}=u^\mu=\left(\dot v, \dot\xi, \dot{\bar\xi}, \dot u\right)
  \ ,  \nonumber\\
&&e^\mu_{(1)}={1\over{\sqrt 2}}\left({2\over{g_{03}\dot u}}\>
 \R e\left\{(g_{12}\dot\xi+g_{22}\dot{\bar\xi}+g_{23}\dot u)\>G_-\right\},
   -\bar G_-, -G_-, 0\right)
  \ ,  \nonumber\\
&&e^\mu_{(2)}={1\over{\sqrt 2}}\left({2\over{g_{03}\dot u}}\>
 \I m\left\{(g_{12}\dot\xi+g_{22}\dot{\bar\xi}+g_{23}\dot u)\>G_+\right\},
   -i\bar G_+, iG_+, 0\right)
  \ , \label{E2.49}\\
&&e^\mu_{(3)}=-\left(\dot v+{1\over{g_{03}\dot u}}, \dot\xi, \dot{\bar\xi},
   \dot u\right)  \ , \nonumber
\end{eqnarray}
where
$\ G_\pm= g_+ \pm g_- \exp(i \hbox{\rm Arg}\> g_{11})$.
The expressions $(\ref{E2.45})$ and $(\ref{E2.49})$ simplify considerably if
$g_{11}=0$ since in this case $g_-=0$ and
$g_+=1/\sqrt{g_{12}}=G_+=G_-$.
\bigskip

\noindent
{\it Proof:} The last equation in (\ref{E2.45}) follows from
Eq. (\ref{E42c}) and $\k\cdot\uu=-1/{\sqrt 2}$,
the equation for $l^\mu$ follows from $\bl=\sqrt 2 \uu - \k$.
Vectors $\m$ and $\bar\m$ can then be determined from
$\e_{\hat a} \cdot \e_{\hat b}=g_{\mu\nu}e^\mu_{\hat a} e^\nu_{\hat b}=
g_{\hat a \hat b}$. The conditions $\m\cdot\k=g_{\hat 1 \hat 4}=
0=g_{\hat 2 \hat 4}=\bar\m\cdot\k$ imply $m^3=0=\bar m^3$,
$\m\cdot\bl=g_{\hat 1 \hat 3}=0$ implies
$m^0=-(l_1m^1+l_2m^2)/l_0$. In given coordinates we have
$\bar m^1=\overline{m^2}$ and $\bar m^2=\overline{m^1}$.
Denoting $X=m^1$ and $Y=m^2$
we get $m^\mu=(-(l_1 X+l_2 Y)/l_0,X,Y,0)$ and
$\bar m^\mu=(-(l_1 \bar Y+l_2 \bar X)/l_0,\bar Y,\bar X,0).$
Functions $X$, $Y$ can be determined as solutions of equations
$\m\cdot\m=g_{\hat 1 \hat 1}=0=g_{\hat 2 \hat 2}=\bar\m\cdot\bar\m$
and $\m\cdot\bar\m=g_{\hat 1 \hat 2}=1$,
\begin{equation}
g_{11}X^2+2g_{12}XY+g_{22}Y^2=0\ , \label{E2.46a}
\end{equation}
\noindent
\begin{equation}
g_{11}X\bar Y+g_{12}(X\bar X+Y\bar Y)+g_{22}\bar X Y=1
\ . \label{E2.46b}
\end{equation}
(i) Assume $g_{11}\not=0.$ Then $X\not=0$, and introducing
a complex function $C$ such that $Y=CX$, Eq. (\ref{E2.46a}) implies
$C=(-g_{12}\pm\sqrt D)/g_{22}$, and Eq. (\ref{E2.46b}) gives
$X\bar X=|X|^2=(g_{12}\pm\sqrt D)/(2D)$. Since $X=|X|e^{i\varphi}$,
$\varphi$ being a real function, we have
$m^1=\sqrt{(g_{12}\pm\sqrt D )/(2D)}\;
  \exp(i(\vartheta-\hbox{Arg}\>g_{11}))$,
$m^2=-\sqrt{(g_{12}\mp\sqrt D )/(2D)}\;\exp(i\vartheta)$,
where $\vartheta=\varphi+\hbox{Arg}\>g_{11}$.
The change from the upper to lower signs
accompanied by $\vartheta\rightarrow -\vartheta+\pi+\hbox{Arg}\>g_{11}$
results just in $\m\leftrightarrow\bar\m$, corresponding to a
reflection $\e_{(2)}\leftrightarrow -\e_{(2)}$.
By performing rotation $m^\mu\rightarrow m'^\mu
=e^{-i\vartheta} m^\mu$ we can write the representatives
of $\m$ and $\bar\m$ given by $\vartheta=0$ so that we arrive at
Eq. (\ref{E2.45}).\\
(ii) If $g_{11}=0$, we simply find  $m^1=0$ and
$m^2=-1/\sqrt{g_{12}}$.
Hence, the null tetrad has the form (\ref{E2.45}),
and this implies the orthonormal frame (\ref{E2.49}).
\bigskip

In general, the frames $\{ \e_a \}$ and $\{ \e_{\hat a} \}$,
related by Eq. $(\ref{E2.9})$, are not parallelly transported
along the geodesic with tangent $\uu=\e_{(0)}$. However,
they are if ${D\k}/{d\tau}=0={D\m}/{d\tau}$.
Starting with an arbitrary $\m$,
the second condition can always be satisfied by choosing
$\m_\parallel=e^{i\vartheta_\|}\,\m$, where
$\vartheta_\| = i\int_0^\tau
\bar\m\cdot({D\m}/{d\tau})\, d\tau +\vartheta_0$,
$\vartheta_0=\hbox{\rm const.}$ We thus arrive at

\bigskip
\noindent
{\it Proposition 3}:
Consider a geodesic $x^\mu(\tau)=(v,\xi,\bar\xi,u)$
in spacetime with metric $(\ref{E2.40})$. Then the orthonormal frame
$\{ \e_a \}$ given by Eq. $(\ref{E2.49})$ and the null tetrad
$\{\e_{\hat a}\}$ given by  Eq. $(\ref{E2.45})$ are parallelly
transported along the geodesic if
\begin{equation}
g_{12,0}\dot \xi+g_{22,0}\dot{\bar\xi}+(g_{23,0}-g_{03,2})\dot u=0
 \ ,\label{E2.51a}
\end{equation}
and
\begin{eqnarray}
\dot\vartheta_\|(\tau)&=&{i\over{2D}}\Big[(G_1\bar G_1-G_2\bar G_2)\> E
  \label{E2.51b}\\
&& + G_1(2D\dot m^1+m^1(g_{12}\dot g_{12}-g_{22}\dot g_{11})+m^2(g_{12}\dot g_{22}-g_{22}\dot g_{12}))
 \nonumber\\
&& + G_2(2D\dot m^2+m^1(g_{12}\dot g_{11}-g_{11}\dot g_{12})+m^2(g_{12}\dot g_{12}-g_{11}\dot g_{12}))\Big]
 \ ,\nonumber
\end{eqnarray}
\noindent
where
$G_1=g_{12}g_-\exp(i \hbox{\rm Arg}\>g_{11})-g_{11}g_+$,
$G_2=g_{22}g_-\exp(i \hbox{\rm Arg}\>g_{11})-g_{11}g_+$,
$E=(g_{12,1}-g_{11,2})\dot\xi+(g_{22,1}-g_{12,2})\dot{\bar\xi}
+(g_{23,1}-g_{13,2})\dot u=-\bar E$,
$m^1=g_- \exp(-\>i \hbox{\rm Arg}\> g_{11})$, and $m^2=-g_+$.
If, in addition,  $g_{11}=0$ then $G_1=0$, $G_2=-\sqrt{g_{12}}$
and Eqs. $(\ref{E2.51a})$, $(\ref{E2.51b})$ reduce to
\begin{equation}
g_{12,0}\dot \xi+(g_{23,0}-g_{03,2})\dot u=0\ ,\label{E2.52a}
\end{equation}
and
\begin{equation}
\dot\vartheta_\|(\tau)=-{i\over2}{1\over{g_{12}}}
 \Big[g_{12,1}\dot \xi-g_{12,2}\dot{\bar\xi}+(g_{23,1}-g_{13,2})
 \dot u\Big]\ .\label{E2.52b}
\end{equation}

\bigskip

\noindent
{\it Proof:}
Using $k_\mu k^\mu=0$ and $k_\mu u^\mu=-1/\sqrt 2$, it can be
shown that $\k$ is parallelly transported if
$\G1_{0\alpha}u^\alpha=0$. Calculating the Christoffel symbols
for the metric (\ref{E2.40}) we find (\ref{E2.51a}).
For proving Eq. (\ref{E2.51b})
we use $m^3=0$ and $\bar m_0=0$, again the condition
$\G1_{0\alpha}u^\alpha=0$ and other Christoffel symbols
for the metric (\ref{E2.40}).

\bigskip
\bigskip
\noindent
{\bf IV. DEVIATION OF GEODESICS IN THE VACUUM NON-TWISTING TYPE {\it N}
   SPACETIMES WITH COSMOLOGICAL CONSTANT}

\bigskip
In this section we apply results given above to the non-twisting
type {\it N} vacuum spacetimes with non-vanishing $\Lambda$.
In the preceding paper \cite{BPI} we showed that all such solutions belong either
to the Kundt class of non-expanding gravitational waves which we
denoted by symbol $KN(\Lambda)$, or to the Robinson-Trautman class of
expanding gravitational waves $RTN(\Lambda,\epsilon)$.

The class $KN(\Lambda)$ can be divided into six invariant
canonical subclasses $KN(\Lambda)[\alpha,\beta]$, and the class
$RTN(\Lambda,\epsilon)$ into nine invariant canonical subclasses,
as analyzed in detail in \cite{BPI}. All
$KN(\Lambda)$ metrics can be written in the form of Eq. (1, I),
all $RTN(\Lambda,\epsilon)$ are described by Eq. (19, I). Both
classes of metrics are of the form (\ref{E2.40}) in coordinates
$x^\mu=(v,\xi,\bar\xi,u)$. In the $KN(\Lambda)$
class we have
\begin{equation}
g_{12}={1\over p^2}\ ,\quad
g_{03}=-{q^2\over p^2}\ ,\quad
g_{33}=F\ ,            \label{E4.2}
\end{equation}
where
$p=1+ \la \xi\bar\xi$,
$q=(1-\la \xi\bar\xi)\alpha+\bar\beta\xi+\beta\bar\xi$,
$F=\kappa(q^2/p^2)v^2 - ((q^2)_{,u}/p^2)v - (q/p)H$,
$\kappa={\Lambda\over 3}\alpha^2+2\beta\bar\beta$,
$H=(f_{,\xi}+\bar f_{,\bar\xi})-(\Lambda/3p)
    (\bar\xi f + \xi\bar f)$;
and in the $RTN(\Lambda,\epsilon)$ class
\begin{equation}
g_{12}=v^2\ ,\quad
g_{13}=v\bar {\Aa}\ ,\quad
g_{23}=v\Aa\ ,\quad
g_{03}=\psi\ ,\quad
g_{33}=2(\Aa\bar {\Aa}+\psi \BB)\ ,    \label{E4.3}
\end{equation}
where $\Aa=\epsilon\xi-v f$,
$\BB=-\epsilon+{v\over2}(f_{,\xi}+\bar f_{,\bar\xi})+\la v^2\psi$,
$\psi=1+\epsilon\xi\bar\xi$, $\epsilon=-1,0,+1$, respectively.

Hence, for the $KN(\Lambda)$ solutions the orthonormal frame
$(\ref{E2.49})$ is given by
\begin{eqnarray}
e^\mu_{(0)}&=&\left(\dot v,\dot\xi,\dot{\bar\xi},\dot u\right)
  \ ,\nonumber\\
e^\mu_{(1)}&=&-{p\over{\sqrt 2}}\left({2\over q^2}{\R e\dot\xi\over\dot u},1,1,0\right)
  \ ,\label{E4.4}\\
e^\mu_{(2)}&=&-{p\over{\sqrt 2}}\left({2\over q^2}{\I m\dot\xi\over\dot u},i,-i,0\right)
  \ ,\nonumber\\
e^\mu_{(3)}&=&-\left(\dot v-{p^2\over\dot u q^2},\dot\xi,\dot{\bar\xi},\dot u\right)
  \ ,\nonumber
\end{eqnarray}
and for the $RTN(\Lambda,\epsilon)$ solutions we have
\begin{eqnarray}
e^\mu_{(0)}&=&\left(\dot v,\dot\xi,\dot{\bar\xi},\dot u\right)
  \ ,\nonumber\\
e^\mu_{(1)}&=&{1\over{\sqrt 2}}{1\over v}\left({2v\over \psi\dot u}
  {\R e\{v\dot\xi+\Aa\dot u\}},-1,-1,0\right)
  \ ,\label{E4.5}\\
e^\mu_{(2)}&=&{1\over{\sqrt 2}}{1\over v}\left({2v\over \psi\dot u}
  {\I m\{v\dot\xi+\Aa\dot u\}},-i,i,0\right)
  \ ,\nonumber\\
e^\mu_{(3)}&=&-\left(\dot v+{1\over\psi\dot u },
  \dot\xi,\dot{\bar\xi},\dot u\right)
  \ .\nonumber
\end{eqnarray}
According to Proposition 3 these frames are parallelly
transported along  time-like geodesics $x^\mu(\tau)=(v,\xi,\bar\xi,u)$
in the $KN(\Lambda)[\alpha, \beta]$ spacetimes if
\begin{equation}
\left( {q\over p} \right)_{,\xi}=0=\left( {q\over p} \right)_{,\bar\xi}
  \ ,\qquad
\dot\vartheta_\|(\tau)=i\left( {{p_{,\xi}}\over p}\dot\xi-
  {{p_{,\bar\xi}}\over p}\dot{\bar\xi}\right)
  \ ,\label{E4.6}
\end{equation}
and in the case of $RTN(\Lambda,\epsilon)$ solutions if
\begin{equation}
\dot\xi=f\dot u\ ,\quad \quad\dot{\bar\xi}=\bar f\dot u
  \ ,\qquad
\dot\vartheta_\|(\tau)={i\over2}\left( f_{,\xi}- \bar f_{,\bar\xi}\right)\dot u
  \ .\label{E4.7}
\end{equation}

The equation of geodesic deviation is now given by Eq. $(\ref{E2.12a})$
with $T_{(a)(b)}=0$. The amplitudes $(\ref{E2.12b})$ for both classes
of spacetimes are calculated in the Appendix. We  find that the
invariant form of the equation of geodesic deviation with respect to
the interpretation frame along any time-like geodesic
in the $KN(\Lambda)$ and $RTN(\Lambda,\epsilon)$ spacetimes takes the form
\begin{eqnarray}
\ddot Z^{(1)}&=&{\Lambda\over3}Z^{(1)}-\A_+ Z^{(1)}+\A_\times Z^{(2)}
   \ ,\nonumber\\
\ddot Z^{(2)}&=&{\Lambda\over3}Z^{(2)}+\A_+Z^{(2)}+\A_\times Z^{(1)}
   \ ,\label{E4.12}\\
\ddot Z^{(3)}&=&{\Lambda\over3}Z^{(3)}\ ,\nonumber
\end{eqnarray}
where the amplitudes of the transverse gravitational wave are given by
\begin{equation}
\A_+(\tau)={\pul} pq\dot u^2\> \R e\left\{f_{,\xi\xi\xi}\right\}\ ,\qquad
\A_\times (\tau)={\pul} pq\dot u^2\> \I m\left\{f_{,\xi\xi\xi}\right\}
   \ , \label{E4.13}
\end{equation}
for the $KN(\Lambda)$  spacetimes, and by
\begin{equation}
\A_+(\tau)=-{\pul} {\psi\over v}\dot u^2\> \R e\left\{f_{,\xi\xi\xi}\right\}
  \ ,\qquad
\A_\times(\tau)=-{\pul} {\psi\over v}\dot u^2\> \I m\left\{f_{,\xi\xi\xi}\right\}
   \ , \label{E6.12}
\end{equation}
in the $RTN(\Lambda,\epsilon)$ spacetimes (see Eqs. (\ref{A11}) and (\ref{A20})
in Appendix). Equations (\ref{E4.12})-(\ref{E6.12}) give relative
accelerations of the free test particles in terms of their actual positions.
They enable us to draw a number of simple conclusions:
\begin{enumerate}

\item  All particles move isotropically one with respect to the other
according to Eqs. (\ref{E2.16c}) if no gravitational wave is present,
i.e., if $f_{,\xi\xi\xi}=0$. In this case both the $KN(\Lambda)$ and
$RTN(\Lambda,\epsilon)$ spacetimes are vacuum conformally flat
(cf. (\ref{A6}), (\ref{A17})), and therefore
Minkowski ($\Lambda=0$), de Sitter ($\Lambda>0$) and
anti-de Sitter ($\Lambda<0$) (see Lemma 1 and 3 in \cite{BPI}).
Such spaces are maximally symmetric, homogeneous, isotropic, and
they represent a natural background for other ``non-trivial''
$KN(\Lambda)$ and $RTN(\Lambda,\epsilon)$ type
{\it N} solutions.

\item If amplitudes $\A_+$ and $\A_\times$ do not vanish
($f_{,\xi\xi\xi}\not=0$), the particles are influenced by the
wave (see Eq. (\ref{E2.17}) and subsequent discussion)
in a similar way as they are affected by a standard gravitational wave
on Minkowski background (cf. \cite{MTW}). However, if $\Lambda\not=0$,
the influence of the  wave adds with the (anti-) de Sitter isotropic
expansion (contraction). This makes plausible our interpretation
of the $KN(\Lambda)$ and $RTN(\Lambda,\epsilon)$ metrics as
{\it exact gravitational waves propagating  on the constant curvature
backgrounds}.

\item The wave propagates in the space-like direction
of $\e_{(3)}$ and has a {\it transverse character} since only motions in the
perpendicular directions of $\e_{(1)}$ and $\e_{(2)}$ are affected. The
propagation direction given by $\e_{(3)}$ coincides with the projection of
the Debever-Penrose vector $\k$ on the hypersurface orthogonal to
the observer's velocity $\uu$ (cf. Proposition 1).

\item There are {\it two polarization modes} of the wave
--- ``+''  and ``$\times$'', $\A_+$ and $\A_\times$ being the
amplitudes. Under rotation (\ref{E2.18}) in the transverse
plane they transform according to Eq. (\ref{E2.19})
so that the helicity of the wave is 2, as with linearized
waves on Minkowski background. For the special choice of the frame given by
$\vartheta(\tau)=\vartheta_+=-\pul pq\dot u^2 \hbox{Arg}\,\{f_{,\xi\xi\xi}\}$
for the $KN(\Lambda)$, and by
$\vartheta_+={1\over4} (\psi/ v)\dot u^2 \hbox{Arg}\,\{f_{,\xi\xi\xi}\}$
for the $RTN(\Lambda,\epsilon)$ spacetimes,
the observer views pure ``+''  polarization, and for
$\vartheta_\times=\vartheta_+ +{\pi\over4}$ --- pure ``$\times$'' polarization.

\item The waves have amplitude
$\A=\pul pq\dot u^2\> \left|f_{,\xi\xi\xi}\right|$ for the $KN(\Lambda)$
class and $\A=\pul (\psi/ v)\dot u^2\> \left|f_{,\xi\xi\xi}\right|$
for $RTN(\Lambda,\epsilon)$;
this is invariant under rotations (\ref{E2.18}). However,
the amplitude changes under Lorentz transformations
to another observer $\uu'$ with a spatial velocity
$\vec v=(v_1,v_2,v_3)$ with respect to  the original observer.
For type {\it N} solutions we get
$\A'=(1-v_3)^2/(1-v_1^2-v_2^2-v_3^2)\, \A$.
By increasing speed in the wave-propagation direction $\e_{(3)}$
($v_1=v_2=0,v_3>0$), he
experiences a weakening of the wave
amplitude by factor $(1-v_3)/(1+v_3)$  ($\A'\rightarrow 0$ as
$v_3\rightarrow 1$) , and by moving in the opposite direction,
an  increase of the amplitude
($\A'\rightarrow \infty$ as $v_3\rightarrow -1$).
By increasing speed in the transverse directions $\e_{(1)}$,
$\e_{(2)}$, ($v_1^2+v_2^2\not=0,v_3=0$), she experiences an
increase  by the factor $1/(1-v_1^2-v_2^2)$.
\end{enumerate}

In general, all $KN(\Lambda)$ spacetimes contain singularities except
for the homogeneous {\it pp}-waves \cite{KSMH} given by $p=1=q$ and $f_{,\xi\xi\xi}
=6c_3(u)$, where  $c_3(u)$ is a finite function of $u$. All other $KN(\Lambda)$
spacetimes are singular at $|\xi|=\infty$ where the amplitudes $\A_+$,
$\A_\times$ diverge. Additional singularities in the
amplitudes may occur if the coefficients $c_n(u),\>n\ge3$, of the analytic
expansion of  function $f(\xi,u)$ are badly behaved at some $u$.

$RTN(\Lambda, \epsilon)$ spacetimes also contain singularities.
The character of the singularities depends on parameter $\epsilon$ and
on the form of the function $f(\xi,u)$. As follows from
Eq. (\ref{E6.12}), there is always a singularity at $v=0$.  Another singularity
is given by $\psi=\infty$ which occurs only for $\epsilon\not=0$ at
$|\xi|=\infty$. There may be singularities for special forms of
$f$, namely if $f_{,\xi\xi\xi}=\infty$.
This occurs at $|\xi|=\infty$ if $f$ contains the terms
$c_n\xi^n,\>n\ge4$. Another type of singularities
may appear if some of the coefficients $c_n(u)$
diverge for some values of $u$. Singularities might be considered as
``sources'' of waves; however, it is far from certain whether
non-singular sources ``covering'' the regions in which singularities occur
can be constructed. The singularities of the
$RTN(\Lambda, \epsilon)$ spacetimes can invariantly be
characterized by the non-vanishing invariant constructed recently
\cite{BPr} from the second derivatives of the Riemann tensor.

Finally, we shall discuss a special class of geodesics
explicitly. Since for
$f=f_c=c_0(u)+c_1(u)\xi+c_2(u)\xi^2$
the metrics represent Minkowski, de Sitter or anti-de Sitter
space  there always exists a transformation of
coordinates which brings $g_{\mu\nu}[f=f_c]$ to $g_{\mu\nu}[f=0]$
(see Lemmas 2 and 4 in \cite{BPI}). It is thus sufficient
to consider only the non-trivial part $f_w\equiv f-f_c$ of function
$f(\xi,u)$. Moreover, one can always rearrange its analytic expansion so that
$f=\sum_{n=0}^\infty c_n(u)\xi^n=\sum_{n=0}^\infty \tilde c_n(u)(\xi-\xi_0)^n
=\sum_{n=3}^\infty \tilde c_n(u)(\xi-\xi_0)^n+\tilde f_c$, $\xi_0$ being
an arbitrary complex constant. Therefore, it is natural to consider
structural functions of the form
\begin{equation}
f_w = c_3(u)(\xi-\xi_0)^3+c_4(u)(\xi-\xi_0)^4+...\ .\label{E6.16}
\end{equation}
Consider a {\it special class of  geodesics}
characterized by $\xi=\xi_0=$const. For the $RTN(\Lambda, \epsilon)$
solutions these are geometrically privileged
since the interpretation frame $(\ref{E4.5})$ is parallelly
propagated along them (Eq. (\ref{E4.7}) is satisfied).
One can also find special geodesics for some subclasses of
$KN(\Lambda)$: $\xi=\xi_0$ for the $PP$ subclass,
$\xi_0=\pm \sqrt{{6\over\Lambda}} $ for  $KN(\Lambda)I$,
and $\xi_0=0$ for $KN(\Lambda^-)II$.
The geodesics $\xi=\xi_0$ have the same
forms as geodesics in the ``background''  since
Christoffel symbols for $f=f_w$ and $\xi=\xi_0$
coincide with those for $f=0$. However, the test particles
feel the tidal forces proportional to $\A_+$ and
$\A_\times$ given by Eqs. (\ref{E4.13}), (\ref{E6.12}).
The amplitudes do not vanish since $f_{w,\xi\xi\xi}=6c_3(u)$ is
non-vanishing.

The timelike geodesics $\xi=\xi_0=$const. in the $RTN(\Lambda>0,
\epsilon)[f_w]$ spacetimes are given by
\begin{equation}
v={\alpha\over{1+\epsilon\xi_0\bar\xi_0}}
  \left( C_1\cosh{\tau\over\alpha}
  +C_2\sinh{\tau\over\alpha} \right)\ ,\qquad
\dot u=-\left( C_1\sinh{\tau\over\alpha}+C_2\cosh{\tau\over\alpha}
   +C_3 \right)^{-1}\ ,\label{E6.29}
\end{equation}
where $\alpha=\sqrt{3/\Lambda}$, $C_1,C_2,C_3$ are real constants
satisfying $C_1^2-C_2^2+C_3^2=2\epsilon$.
The integration of Eq. (\ref{E6.29}) can be performed explicitly
but we do not give it here since only $\dot u$ enters the
amplitudes. The wave amplitudes (\ref{E6.12}) are
$\A_+=\R e\,\A$ and $\A_\times=\I m\,\A$ where
\begin{eqnarray}
\A(\tau)&=&-{3\over\alpha} (1+\epsilon\xi_0\bar\xi_0)^2
 \left( C_1\cosh{\tau\over\alpha}+C_2\sinh{\tau\over\alpha} \right)^{-1}
 \times\nonumber\\
&&\hskip18mm \times\left( C_1\sinh{\tau\over\alpha}+C_2\cosh{\tau\over\alpha}+C_3 \right)^{-2}
\> c_3(u(\tau))\ .\label{E6.30}
\end{eqnarray}
As proper time $\tau$ along geodesics increases,
$\tau\to\infty$, particles recede from $v=0$ and amplitudes
decay as $ \A\sim \exp(-3{\sqrt{\Lambda/3}}\,\tau)$,
i.e., {\it waves are damped exponentially}.
The spacetime locally approaches the de Sitter universe. This is
an explicit demonstration of the {\it cosmic no-hair  conjecture}
(see, e.g. \cite{Mae}) under the presence of
waves within exact model spacetimes. (For cosmic
no-hair conjecture in the Robinson-Trautman spacetimes of Petrov
type {\it II} see \cite{BP1}, \cite{BP2}.)

Similarly, for the $KN(\Lambda>0)I[f_w]$
subclass (representing the only spacetimes of the $KN(\Lambda)$
type admitting $\Lambda>0$) the geodesics
$\xi=\xi_0=\pm \sqrt{{6/\Lambda}}$ are given by
\begin{equation}
v=C_1 \exp\>\left(\frac{\tau}{\alpha}\right)\ ,\qquad
u=-{1\over 2C_1} \exp\>\left(-\frac{\tau}{\alpha}\right) +C_2\ ,
\label{E4.33a}
\end{equation}
and
\begin{equation}
v=C_1 \sinh\left(\frac{\tau}{\alpha}+2\tau_0\right)\ ,\qquad
u={1\over 2C_1} \tanh\left(\frac{\tau}{2\alpha}+\tau_0\right) +C_2\ , \label{E4.33b}
\end{equation}
\noindent
with $C_1$, $C_2$, $\tau_0$ constants.
For observers moving along these geodesics,
\begin{equation}
\A(\tau)=\pm 12\sqrt{6\over\Lambda} \dot u^2(\tau)\> c_3(u(\tau))
\ .\label{E4.39}
\end{equation}
After substitution of the explicit dependence of $u(\tau)$ we see
that as $\tau\to +\infty$ the amplitudes behave like
$\A\sim\exp({-2\sqrt{\Lambda/3}\,\tau})$.
Again, gravitational waves are damped exponentially and the
cosmic no-hair conjecture is confirmed.

\bigskip
\bigskip
\noindent
{\bf ACKNOWLEDGMENTS}
\bigskip

We thank Jerry Griffiths for reading the manuscript and for very
helpful comments.
We also acknowledge the support of grant No. GACR-202/99/0261
from the Czech Republic.

\bigskip
\bigskip
\noindent
{\bf APPENDIX: GRAVITATIONAL WAVE AMPLITUDES}

\bigskip
We calculate  amplitudes $\A_+=\pul\R e\Psi_4$ and
$\A_\times=\pul\I m\Psi_4$ by using differential forms.
Let $\{\e_{\hat a}\}=\{\m,\bar{\m},\bl,\k\}$
be a null tetrad, $\m=\e_{\hat 1}=m^\mu\partial_\mu$,
$\bar\m=\e_{\hat 2}=\bar m^\mu\partial_\mu$,
$\bl=\e_{\hat 3}=l^\mu\partial_\mu$,  $\k=\e_{\hat 4}=k^\mu\partial_\mu$.
The dual basis $\{{\bomega}^{\hat a}\}$ is given by one-forms
$\bomega^{\hat 1}=\bar m_\mu dx^\mu$, $\bomega^{\hat 2}=m_\mu dx^\mu$,
$\bomega^{\hat 3}=-k_\mu dx^\mu$, $\bomega^{\hat 4}=-l_\mu
dx^\mu$; the metric is $ds^2=g_{\hat a\hat b}\>\bomega^{\hat a}\bomega^{\hat b}=
2\>\bomega^{\hat 1}\bomega^{\hat 2}-2\>\bomega^{\hat 3}\bomega^{\hat 4}$
with $g_{\hat1\hat2}=\e_{\hat1}\cdot\e_{\hat2}=1$ and
$g_{\hat3\hat4}=\e_{\hat3}\cdot\e_{\hat4}=-1.$
The natural choice of the null basis for the metric $KN(\Lambda)$ is
\begin{equation}
  \bomega^{\hat 1}={d\bar\xi\over p}\ ,\quad
  \bomega^{\hat 2}={d\xi\over p}    \ ,\quad
  \bomega^{\hat 3}={q^2\over p^2}du \ ,\quad
  \bomega^{\hat 4}=dv-{1\over 2}{p^2\over q^2}Fdu \ ;\label{A4}
\end{equation}
in coordinates $x^\mu=(v,\xi,\bar\xi,u)$ we have
\begin{eqnarray}
&&m^\mu=(0,0,p,0)   \ ,\qquad
 {\bar m}^\mu=(0,p,0,0)   \ ,  \label{A5}\\
&&k^\mu=(1,0,0,0)\ ,\qquad
 l^\mu=({1\over 2}{p^4 \over q^4}F,0,0,{p^2\over q^2})
  \ .\nonumber
\end{eqnarray}
The non-vanishing components of the Weyl tensor in this
null tetrad are
\begin{equation}
C_{\hat 3\hat 2\hat 3\hat 2}\equiv\Psi_4=
 {1\over2}{p^5\over q^3}f_{,\xi\xi\xi}= \overline{C_{\hat 3\hat 1\hat 3\hat 1}}
  \ .  \label{A6}
\end{equation}
The interpretation null tetrad for the $KN(\Lambda)$ spacetimes,
given by  Eq. (\ref{E2.45}), reads
\begin{eqnarray}
&&m^\mu=\Bigg( -{p\over q^2}{\dot\xi\over\dot u},0,-p,0 \Bigg)
  \ ,\qquad
{\bar m}^\mu=\Bigg(-{p\over q^2}{\dot{\bar\xi}\over\dot u},-p,0,0\Bigg)
  \ ,  \label{A8}\\
&&k^\mu=\left( {1\over{\sqrt 2 \dot u}}{p^2\over q^2} , 0 , 0 , 0
  \right)\ ,\qquad
l^\mu=\left( \sqrt 2 \dot v-{1\over{\sqrt 2 \dot u}}{p^2\over q^2} ,
              \sqrt 2 \dot\xi , \sqrt 2 \dot{\bar\xi} ,  \sqrt 2 \dot u \right)
  \ .\nonumber
\end{eqnarray}
The relation between tetrads (\ref{A5}) and (\ref{A8})
is given by the Lorentz transformation,
$\k_{natur}=A\k_{interp}$,
$\bl_{natur}=(\bl_{interp}+Be^{i\vartheta}\bar\m_{interp}
 +\bar B e^{-i\vartheta}\m_{interp}+B\bar B\> \k_{interp})/A$,
$\m_{natur}=e^{-i\vartheta}\m_{interp}+B\>\k_{interp}$,
where
$A=\sqrt 2 \dot u\, q^2/p^2$,
$B=-\sqrt2 \,\dot\xi/ p$,
$\vartheta=\pi$.
The coefficients $\Psi_A$ transform (see \cite{KSMH})
\begin{eqnarray}
&&\Psi_4^{interp}=A^2\Psi_4^{natur}=pq\dot u^2 f_{,\xi\xi\xi}
  \ ,  \label{A11}\\
&&\Psi_3^{interp}=\Psi_2^{interp}=\Psi_1^{interp}=\Psi_0^{interp}=0
  \ . \nonumber
\end{eqnarray}

Similarly, the natural choice of null basis for the $RTN(\Lambda, \epsilon)$
metric is
\begin{equation}
\bomega^{\hat 1}=vd\bar\xi+\bar {\Aa} du\ ,\quad
\bomega^{\hat 2}=vd\xi+\Aa du\ ,\quad
\bomega^{\hat 3}=\psi du\ ,\quad
\bomega^{\hat 4}=-dv-\BB du \ ,\label{A15}
\end{equation}
so that
\begin{eqnarray}
&&m^\mu=(0,0,{1\over v},0) \ ,\qquad
 \bar m^\mu=(0,{1\over v},0,0)  \ ,\label{A16}\\
&&k^\mu=(-1,0,0,0)\ ,\qquad
 l^\mu=(-{\BB\over\psi},-{\Aa\over{v\psi}},
 -{\bar {\Aa}\over{v\psi}},{1\over \psi})\ .\nonumber
\end{eqnarray}
For this tetrad we obtain non-vanishing components
\begin{equation}
C_{\hat 3\hat 2\hat 3\hat 2}\equiv\Psi_4=
 -{1\over{2v\psi}}\>f_{,\xi\xi\xi}= \overline{C_{\hat 3\hat 1\hat 3\hat 1}}
  \ .  \label{A17}
\end{equation}
The interpretation null tetrad (\ref{E2.45}) reads
\begin{eqnarray}
&&m^\mu=\Bigg( {1\over \psi\dot u}(v\dot\xi+\Aa \dot u),0,-{1\over v},0 \Bigg)
  \ ,\qquad
{\bar m}^\mu=\Bigg({1\over \psi\dot u}(v\dot{\bar\xi}+\bar {\Aa}\dot u),-{1\over v},0,0\Bigg)
  \ ,  \label{A18}\\
&&k^\mu=\left(- {1\over{\sqrt 2 }}{1\over \psi\dot u}, 0, 0, 0
  \right)\ ,\qquad
 l^\mu=\left( \sqrt 2 \dot v+{1\over{\sqrt 2 }}{1\over \psi\dot u} ,
              \sqrt 2 \dot\xi , \sqrt 2 \dot{\bar\xi} ,  \sqrt 2 \dot u\right)
  \ .\nonumber
\end{eqnarray}
The relation between the tetrads (\ref{A16}) and  (\ref{A18}) is
again given by the Lorentz transformation with
$A=\sqrt 2\, \dot u \psi$,
$B=-\sqrt 2\, (v\dot\xi+\Aa\dot u)$,
$\vartheta=\pi$.
We thus get
\begin{eqnarray}
&&\Psi_4^{interp}=A^2\Psi_4^{natur}=-{\psi\over v}\dot u^2\>f_{,\xi\xi\xi}
  \ ,  \label{A20}\\
&&\Psi_3^{interp}=\Psi_2^{interp}=\Psi_1^{interp}=\Psi_0^{interp}=0
  \ . \nonumber
\end{eqnarray}

\end{document}